\documentclass[aps,prd,twocolumn,floatfix,nofootinbib,superscriptaddress]{revtex4-1}
\usepackage{dcolumn}

\usepackage{amsmath,amsfonts,extarrows,amssymb,mathrsfs,times,bm}
\usepackage{extarrows}
\usepackage{graphicx}
\usepackage{float}
\usepackage{graphicx,epstopdf}
\usepackage{dcolumn}
\usepackage{exscale}
\usepackage{relsize}
\usepackage{mathtools}
\usepackage{mhchem}
\usepackage{enumitem}
\usepackage{physics}
\usepackage[colorlinks=true,breaklinks=true,linkcolor=blue,citecolor=blue,urlcolor=blue]{hyperref}
\newcommand{\nn}{\nonumber}

\begin{document}

\title{{\LARGE From Spray to Metric}\\[4pt]
	{\large The Geometric Construction of the Jacobi Metric}}

\author{Zonghai Li}
\email[Electronic address:~]{lizzds@gznu.edu.cn}
\affiliation{School of Physics and Electronic Science, Guizhou Normal University, Guiyang, 550025, China}
\date{\today}
\begin{abstract}
This paper develops a systematic approach to the geometrization of dynamics from the viewpoint of the geodesic equation. The method promotes a semispray to a spray through the imposition of suitable dynamical constraints, and the associated metric structure is extracted via reparameterization. When applied to static spacetimes, this spray-to-metric framework recovers the optical metric, the Jacobi metric for massive particles, and its generalization for charged particles in electromagnetic fields. We further show that a Randers-type Finsler metric arises naturally in the planar circular restricted three-body problem. By establishing a direct pathway from equations of motion to metric structures, this work offers a geometric perspective, independent of the traditional variational framework, may provide a basis for further studies on dynamical systems.
\end{abstract}
\maketitle


\section{Introduction}

A \emph{spray} on a smooth manifold $M$ is a vector field $G$ on the slit tangent bundle $TM\setminus\{0\}$ that is homogeneous of degree two in the fiber coordinates~\cite{Shen_Spray}. The projections of its integral curves onto the manifold are called the geodesics of the spray. Every Riemannian metric, and more generally any Finsler metric, naturally induces such a spray, whose geodesics coincide with those of the metric~\cite{BCS_Finsler}. However, the converse is not true: not every spray necessarily arises from a metric. A central question in spray geometry is to determine when a given spray admits a metric structure—this is known as the metrizability problem for sprays~\cite{Matsumoto1995,Li_Shen2018,Li_Mo_Yu2019}. Relaxing the condition of quadratic homogeneity yields a \emph{semispray}, whose geodesics correspond to general systems of second-order ordinary differential equations (SODEs).

Along the line of second-order ordinary differential equations, a natural hierarchy emerges linking geometry and dynamics—from metrics to sprays, and further to semisprays. Given the intuitive insight and explanatory power of geometry, it is natural to seek a geometric approach to dynamical systems—that is, to pursue the geometrization of dynamics. One may then ask whether an arbitrary system of SODEs can possess geometric structures analogous to those defined by a metric. This question is addressed within the geometric framework known as the Kosambi–Cartan–Chern (KCC) theory~\cite{Antonelli-2003}. The essence of KCC theory is to equip a semispray with a nonlinear connection and curvature-like geometric invariants. This machinery captures the dynamics and stability of systems in purely geometric terms, with applications extending across multiple disciplines~\cite{Sabau2005NA,Sabau2005RWA,Harko2015}.

Although KCC theory provides a powerful geometric framework with broad applicability, its intrinsically non-metric nature limits the use of many sophisticated techniques developed in metric geometry. This limitation naturally suggests the development of complementary approaches that incorporate metric structures into the geometric analysis of dynamical systems.

A classical approach to the metric geometrization of dynamical systems is provided by the Jacobi metric. Jacobi originally established this framework by giving a geometric formulation of the Maupertuis principle of least action, demonstrating that for energy-conserving systems, the trajectories of massive particles coincide with geodesics of the Jacobi metric~\cite{Arnold-mech,Awrejcewicz-mech}. This formulation has been extensively applied in Newtonian gravity~\cite{Pin1975,Szydlowski1996,Chanda-IJGMMP-2017}. Recently, the framework has been extended to curved spacetime. Gibbons derived the Jacobi metric for neutral particles in static spacetimes~\cite{Gibbons2016}, and Chanda \emph{et~al.} subsequently generalized it to stationary spacetimes~\cite{Chanda2019,Chanda_hsdf}. The Jacobi metric describes the case of massive particles, while for massless particles the corresponding framework is provided by the optical metric, derived from Fermat's principle~\cite{Perlick}. In curved spacetime, both the Jacobi and optical metrics, which may take Riemannian or more general Randers–Finsler forms, provide powerful differential-geometric tools for studying phenomena such as particle dynamics and gravitational lensing.
~\cite{Gibbons-Werner,Werner2012,ISOA2016,Crisnejo-Gallo,massiveGB-CGJ,massiveGB-LiWJ,massiveGB-LiJa2020,Qiao-PRD-2022b,Qiao-EPJC-2025,Cunha-CQG-2022,massiveGB-LiHZ,HuangCa,Arganaraz-CQG-2021,Das-EPJC-2017}. Compared with the KCC theory, which provides a non-metric geometric reformulation of dynamics, the Jacobi metric method yields a richer metric–geometric framework, but its reliance on variational principles and energy conservation limits its applicability.

A comparison between the KCC theory and the Jacobi metric method reveals a tension between generality and geometric richness, arising partly from the intrinsic asymmetry in the hierarchy from metrics to sprays and semisprays, and partly from the limitations of existing approaches. Motivated by this, we seek a way to reverse this one-way correspondence by introducing suitable dynamical constraints that promote a semispray to a spray and subsequently to a metric structure. In this framework, metric reconstruction proceeds directly from the SODEs, thereby preserving geometric richness while extending applicability. As a first step, we test the method in three classical cases in static spacetimes—the optical metric, the Jacobi metric for massive particles, and its charged-particle extension—and further illustrate its utility in the planar circular restricted three-body problem.

The paper is organized as follows. In Sec.~\ref{Newton}, we introduce the concepts of sprays and semisprays, together with the spray induced by a Finsler metric. We then describe our method for extracting a metric by promoting a semispray to a spray. In addition, as preparation for later developments, we derive the equation of 3-acceleration for a particle in a general four-dimensional curved spacetime. In Secs.~\ref{Hamilton}, \ref{Einstein}, and \ref{Maxwell}, we apply this method in static spacetimes to reconstruct the optical metric, the Jacobi metric for massive particles, and the Jacobi metric for charged particles in electromagnetic fields, respectively. Sec.~\ref{Poincaré} is devoted to the application of the approach to the planar circular restricted three-body problem. We conclude with a summary in Sec.~\ref{Li Zonghai}. Throughout this work, we use geometrized units, in which the speed of light and Newton’s gravitational constant are set to unity ($c=G=1$).


\section{Geometric Preliminaries and Method}
\label{Newton}

\subsection{Spray}

Let $M$ be a smooth $n$-dimensional manifold and $TM$ its tangent bundle.  
A \emph{spray} on $M$ is a smooth vector field $G$ defined on the slit tangent bundle $TM\setminus\{0\}$, which in local coordinates $(x^i,y^i)$ is written as~\cite{Shen_Spray}
\begin{equation}
	G \;=\; y^i \frac{\partial}{\partial x^i} \;-\; 2\,G^i(x,y)\,\frac{\partial}{\partial y^i}.
\end{equation}
The functions $G^i(x,y)$, called the spray coefficients, are required to be positively homogeneous of degree two in $y$, that is,
\begin{equation}
	\label{Borges}
	G^i(x,\lambda y) \;=\; \lambda^{2} G^i(x,y), \qquad \lambda>0.
\end{equation}
A manifold equipped with a spray is called a spray space. 

The geometric interpretation of a spray is closely tied to the theory of second-order ordinary differential equations.  
Indeed, the integral curves of $G$ project to curves $\gamma(t)$ on $M$ that satisfy
\begin{equation}
	\frac{d^2\gamma^i}{dt^2} + 2G^i\!\left(\gamma, \frac{d\gamma}{dt}\right) = 0 ,
\end{equation}
which are precisely the geodesics of the spray $G$.  
If the homogeneity condition~\eqref{Borges} is not satisfied, $G$ is called a \emph{semispray}.  
In this case, the geodesic equations of the semispray, regarded as a system of second-order ordinary differential equations, form the fundamental object of study in KCC (Kosambi–Cartan–Chern) theory~\cite{Antonelli-2003}.

Every Finsler metric induces a spray. A Finsler structure $F: TM \to [0, \infty)$ is defined as a function that is smooth on $TM \setminus \{0\}$, positively homogeneous of degree one in $y$, i.e.,
\begin{equation}
	F(x,\lambda y) = \lambda F(x,y), \quad \lambda>0,
\end{equation}
and for which the fundamental tensor
\begin{equation}
	g_{ij}(x,y) = \tfrac{1}{2}\,\frac{\partial^2 F^2}{\partial y^i \partial y^j}
\end{equation}
forms a positive definite matrix for all $(x,y)\in TM\setminus\{0\}$.

The spray coefficients associated with $F$ are given by~\cite{Cheng-Shen}
\begin{equation}
	G^i(x,y) = \tfrac{1}{4}\, g^{il}(x,y)\left(
	\frac{\partial^2 F^2}{\partial x^k \partial y^l} y^k
	- \frac{\partial F^2}{\partial x^l}
	\right).
\end{equation}
This spray of the Finsler manifold $(M,F)$ generates the geodesics of $F$.  
In what follows, we present the sprays corresponding to two widely studied Finsler metrics, the Riemannian metric and the Randers metric, which are of particular significance in the present study.

\subsubsection{Riemannian Spray}
\label{sssec:riemannian-spray}

Riemannian geometry is the quadratic special case of Finsler geometry, where the Finsler function is
\begin{equation}
	F(x,y) = \sqrt{g_{ij}(x)\,y^i y^j}.
\end{equation}
In this situation, the fundamental tensor reduces to $g_{ij}(x,y)=g_{ij}(x)$, i.e.\ it is independent of $y$.  
The spray coefficients for a Riemannian metric are
\begin{equation}
	G^i(x,y) = \tfrac{1}{2}\,\Gamma^i_{jk}(x)\,y^j y^k,
\end{equation}
where $\Gamma^i_{jk}$ denote the Christoffel symbols,
\begin{equation}
	\Gamma^i_{jk} = \tfrac{1}{2}\, g^{il}\,
	\bigl(\partial_j g_{kl} + \partial_k g_{jl} - \partial_l g_{jk}\bigr).
\end{equation}
We note that the spray coefficients of a Riemannian metric are quadratic in $y$; however, not all sprays with this property are Riemannian.  Sprays whose coefficients are quadratic in $y$ are referred to as Berwald sprays.

For a Riemannian metric, the associated spray yields the familiar geodesic equation
\begin{equation}
	\frac{d^2x^i}{dt^2} + \Gamma^i_{jk}\,\frac{dx^j}{dt}\frac{dx^k}{dt} = 0.
\end{equation}

Now consider two conformally related metrics $g=g_{ij}y^iy^j$ and $h=h_{ij}y^iy^j$, with
\begin{equation}
	\label{Zagajewski}
	h= \Phi(x)\, g, \quad \Phi(x)>0.
\end{equation}
The Christoffel symbols of $h$, denoted $\tilde{\Gamma}^i{}_{jk}$, are related to those of $g$ by
\begin{equation}
	\label{Ernaux}
	\tilde{\Gamma}^i{}_{jk}
	= \Gamma^i{}_{jk}
	+ \tfrac{1}{2}\Bigl(
	\delta^i_j\,\partial_k \ln \Phi
	+ \delta^i_k\,\partial_j \ln \Phi
	- g_{jk}\, g^{il}\partial_l \ln \Phi
	\Bigr).
\end{equation}

Consequently, the sprays $G^i$ and $\tilde{G}^i$ are related by
\begin{equation}
	\label{Perec}
	\tilde{G}^i
	= G^i
	+ \tfrac{1}{4}\Bigl(
	\delta^i_j\,\partial_k \ln \Phi
	+ \delta^i_k\,\partial_j \ln \Phi
	- g_{jk}\, g^{il}\partial_l \ln \Phi
	\Bigr)y^j y^k.
\end{equation}
This makes explicit how the spray transforms under a conformal change of the Riemannian metric.

\subsubsection{Randers Spray}

A Randers metric is a Finsler metric of the form~\cite{Randers}
\begin{equation}
	F(x,y) = \alpha(x,y) + \beta(x,y),
\end{equation}
where $\alpha(x,y) = \sqrt{a_{ij}(x)y^i y^j}$ is the Riemannian part and $\beta(x,y) = b_i(x)y^i$ is a 1-form, with the requirement
\begin{equation}
	\|\beta\|_\alpha^2 = a^{ij}(x)b_i(x)b_j(x) < 1 ,
\end{equation}
which ensures that $F$ is a Finsler metric.

The spray coefficients of a Randers metric are given by~\cite{Cheng-Shen}
\begin{equation}
	\label{chern}
	G^i \;=\; G^i_{\alpha} \;+\; \tfrac{1}{2} l^i r_{jk} y^j y^k 
	\;+\; \bigl(a^{ij} - l^i b^j\bigr) s_{jk}\alpha y^k,
\end{equation}
where $G^i_{\alpha}$ are the spray coefficients of the Riemannian metric $\alpha$, and
\begin{align}
	\begin{split}
		l^i &:= \frac{y^i}{F}, \quad r_{ij} := \tfrac{1}{2}(b_{i|j}+b_{j|i}), \\
		s_{ij} &:= \tfrac{1}{2}(b_{i|j}-b_{j|i})=\tfrac{1}{2}(b_{i,j}-b_{j,i}). 
	\end{split}
\end{align}
Here $b_{i|j}$ denotes the covariant derivative of $b_i$ with respect to the Levi–Civita connection of $\alpha$. We use both the comma “,” notation and the symbol $\partial$ to indicate partial derivatives, i.e. $b_{i,j} \equiv \partial_j b_i$. It should be noted that, the Randers metric is Berwald if and only if the $1$-form $\beta$ is parallel with respect to $\alpha$, that is, $b_{i|j} = 0$.

The explicit expression of the Randers spray~\eqref{chern} is rather involved.  
A convenient simplification is obtained by reparametrizing the geodesics so that the Riemannian part $\alpha$ has constant speed, $T^i=\tfrac{dx^i}{ds}$.  
In this parametrization, the geodesic equations reduce to (see Ref.~\cite{BCS_Finsler}, p.~297, Exercise~11.3.3)
\begin{equation}
	\frac{d^2 x^i}{ds^2} 
	+ \gamma^i_{jk}\,\frac{dx^j}{ds}\frac{dx^k}{ds}
	+ a^{ij}\left(b_{j,k}-b_{k,j}\right)\,\alpha(T)\,\frac{dx^k}{ds} = 0,
\end{equation}
where $\gamma^i_{jk}$ are the Christoffel symbols of the Riemannian metric $\alpha$ and 
$\alpha(T)=\sqrt{a_{mn}T^mT^n}$ denotes the $\alpha$-speed of the curve.

Equivalently, the corresponding spray coefficients can be expressed as
\begin{equation}
	\label{Russell}
	G^i \;=\; G^i_{\alpha} \;+\;\tfrac12\,a^{ij}\left(b_{j,k}-b_{k,j}\right)\,\alpha(y)\,y^k.
\end{equation}

\subsection{The Method}

Every Finsler metric induces a spray, but the converse is not true. For an arbitrary spray, the existence of a corresponding metric is an interesting and important mathematical problem, closely related to the geometrization of dynamical systems. Rather than attempting a general proof, we focus on certain special dynamical systems and demonstrate how they can be geometrized. Typically, a dynamical system is associated with a \emph{semispray}. Our basic idea can be summarized as follows: starting from a semispray, we employ the intrinsic constraints of the system to rewrite it as a spray, and then, through a suitable reparametrization and by exploiting conformal relations, we identify the metric associated with the spray:
\begin{equation}
	\fbox{$\text{semispray} \;\;\longrightarrow\;\; \text{spray} \;\;\longrightarrow\;\; \text{metric}$}
\end{equation}

To illustrate the method, let us consider the Newtonian dynamics of a particle of mass $m$ moving in a conservative potential $U(x)$. The equation of motion is
\begin{equation}
	m\,\ddot{x}^i = -\,\partial^i U(x),
\end{equation}
where $\dot{x}^i = dx^i/dt$ denotes the velocity. This equation can be cast in the form
\begin{equation}
	\ddot{x}^i + 2G^i = 0, 
	\quad 
	G^i(x,\dot{x}) = \frac{1}{2m}\,\partial^i U(x),
\end{equation}
which defines a semispray. At this stage the force term depends only on the coordinates and is not quadratic in the velocities, hence the structure is not yet a genuine spray. 

\subsubsection{From Semispray to Spray}  

In order to obtain a genuine spray, one needs to make use of the energy constraint so that the dynamics can be rewritten in a homogeneous quadratic form. The conserved energy is  
\begin{equation}
	E = \tfrac{1}{2} m\, \delta_{ij}\,\dot{x}^i \dot{x}^j + U(x),
\end{equation}
which allows the velocity magnitude to be eliminated as
\begin{equation}
	\label{Kopernik}
	\delta(x,\dot{x})=\delta_{ij}\,\dot{x}^i \dot{x}^j = \tfrac{2}{m}\,\big(E-U(x)\big).
\end{equation}
With this substitution, the force term can be rewritten as
\begin{align}
	\partial^{i}U(x)\;&\longrightarrow\;
	\partial^{i}U(x)\,\frac{m\,\delta(x,\dot{x})}{2\big(E-U(x)\big)}
	\nonumber\\[2mm]
	&= -\,\frac12\,m\,\partial^{i}\!\ln\!\Phi(x)\; \delta(x,\dot{x}),
\end{align}
where we have defined
\begin{equation}
	\Phi(x) = E - U(x).
\end{equation}
Substituting this relation back into the dynamical equation yields
\begin{equation}
	\ddot{x}^i-\frac12\,\partial^{i}\!\ln\!\Phi(x)\, \delta(x,\dot{x})=0,
\end{equation}
which is homogeneous and quadratic in the velocities. This structure corresponds to a true \emph{spray}, with spray coefficients
\begin{equation}
	G^i(x,\dot x) = -\frac14\,\partial^{i}\!\ln\!\Phi(x)\;\delta(x,\dot{x}).
\end{equation}

\subsubsection{From Spray to Metric}  

Introducing a new parameter $\ell$ along the trajectory, we define
\begin{equation}
	y^i = \frac{dx^i}{d\ell}, 
	\qquad 
	f = \frac{dt}{d\ell}.
\end{equation}
The equation of motion then takes the form
\begin{equation}
	\frac{d^2x^i}{d\ell^2}
	-\frac12\,\partial^{i}\!\ln\!\Phi(x)\;\delta(x,y)
	= \left(\partial_{k}\ln f \right)\,y^k y^i .
\end{equation}
By adding the symmetrized term to both sides of the equation,  
\begin{equation}
	\frac12 \Big[\delta^i{}_j\,\partial_k\!\ln \Phi(x)
	+\delta^i{}_k\,\partial_j\!\ln \Phi(x)\Big] y^j y^k ,
\end{equation}
we obtain
\begin{equation}
	\label{Pauli}
	\frac{d^2x^i}{d\ell^2}
	+ \tilde G^i(x,y)
	= \Big[\partial_{k}\ln(f\Phi)\Big]\,y^k y^i ,
\end{equation}
with
\begin{align}
	\label{Aristotle}
	\tilde G^i(x,y)
	=&\frac14\Big(\delta^i{}_j\,\partial_k\!\ln \Phi(x)
	+\delta^i{}_k\,\partial_j\!\ln \Phi(x)\nn\\
	&
	-\delta_{jk}\,\delta^{il}\,\partial_l\!\ln \Phi(x)\Big)y^j y^k .
\end{align}
Comparing with Eq.~\eqref{Perec}, it is straightforward to verify that these coefficients coincide with the spray of the Riemannian metric
\begin{equation}
	h(x,y)= \Phi(x)\,\delta(x,y)= (E-U(x))\,\delta(x,y).
\end{equation}

One can choose
\begin{equation}
	f\Phi = C = \mathrm{const}, 
	\quad \text{i.e.,}\quad 
	f=\frac{dt}{d\ell}=\frac{C}{E-U(x)},
\end{equation}
to remove the right-hand side of Eq.~\eqref{Pauli} and reduce the trajectories precisely to the geodesics of the metric $h$.  
Up to the conventional constant factor $2m$, the metric $h$ is equivalent to the Jacobi metric,
\begin{equation}
	J=J_{ij}y^iy^j=2m\,(E-U)\,\delta_{ij}y^iy^j.
\end{equation}
Indeed, Eq.~\eqref{Aristotle} shows that replacing $\Phi(x)$ by $C_0\Phi(x)$ with any constant $C_0>0$ yields an equivalent solution, so $h_{ij}$ and $J_{ij}$ differ only by the constant factor $2m$.

With the normalization constant fixed as $C=\sqrt{m/2}$, the dynamics of a particle with fixed energy $E$ in a conservative potential $U(x)$ can thus be described geometrically by
\begin{equation}
	d\ell^2 = (E-U(x))\,\delta_{ij}\,dx^i dx^j,
\end{equation}
valid in the energetically allowed region $E > U(x)$.

In this way, we reinterpret the ``external force'' as an intrinsic geometric effect, thereby achieving a geometrization of the dynamics. From this simple example we observe that the key lies in exploiting energy conservation, which formally leads to a constraint of the type
\begin{equation}
	\label{Yourcenar}
	g(x,y)=g_{ij}(x)\,y^i y^j = \mathcal{F}^2(x),
\end{equation}
a relation we shall refer to as the \emph{quadratic constraint}.  

More generally, the passage from a semispray to a spray can be achieved 
by imposing a constraint that is homogeneous of degree one in $y$. 
A sufficiently general form of such a constraint is
\begin{equation}
	\label{Plato}
	\kappa(x, y) = \mathcal{F}(x),
\end{equation}
where $\kappa(x, y)$ is positively homogeneous of degree one 
in the fiber variable $y$, that is, $\kappa(x, \lambda y) = \lambda\,\kappa(x, y)$ for all $\lambda > 0$. 
The quadratic constraint arises as a special case, in which the 1-homogeneous function 
$\kappa$ is taken to be the norm induced by the metric, 
$\kappa(x, y) = \sqrt{g_{ij}(x)\,y^i y^j}$.

Sprays arising from such generalized constraints~\eqref{Plato} may in fact correspond to Finsler metrics beyond the Riemannian and Randers classes, which are considerably more intricate. In the present work, however, we restrict attention to the quadratic constraint~\eqref{Yourcenar}. In particular, the trajectories of particles moving in a static spacetime satisfy precisely this type of constraint, thereby allowing a geometric reformulation of the dynamics. In what follows, we analyze this case in detail. To begin, let us present the equation of 3-acceleration.

\subsection{Equations for the 3-Acceleration in 4D Spacetime}

\subsubsection{General case}
In coordinates \( (x^0, x^1, x^2, x^3) = (t, x^1, x^2, x^3) \), a stationary metric takes the form
\begin{equation}
	ds^2 = g_{00}(x)\, dt^2 + 2 g_{0i}(x)\, dt\,dx^i + g_{ij}(x)\, dx^i dx^j.
\end{equation}
In the presence of an electromagnetic potential \(A=(A_0,A_i)\), the motion of a charged particle with mass \(m\) and charge \(q\) is described by the four-dimensional Lorentz-force equation parameterized by the affine parameter \(\tau\):
\begin{align}
	\label{LorentzAffine}
	&\frac{d^2 x^\mu}{d\tau^2} + \Gamma^\mu_{\nu\lambda}\,
	\frac{dx^\nu}{d\tau}\frac{dx^\lambda}{d\tau}
	= \frac{q}{m}\,F^\mu{}_{\ \nu}\,\frac{dx^\nu}{d\tau},
	\nn\\
	&~~~~	F_{\mu\nu}=A_{\nu,\mu}-A_{\mu,\nu}.
\end{align}

To express the dynamics in terms of the coordinate time $t$, we note
\begin{align}
	\frac{d^2 x^i}{dt^2} 
	&= \left(\frac{dt}{d\tau}\right)^{-2}\frac{d^2 x^i}{d\tau^2}
	-\left(\frac{dt}{d\tau}\right)^{-3}\frac{d^2 t}{d\tau^2}\frac{dx^i}{d\tau}.
\end{align}

The $i$- and $t$-components of Eq.~\eqref{LorentzAffine} read
\begin{align}
\begin{split}
	\frac{d^2 x^i}{d\tau^2} &= -\Gamma^i_{\nu\lambda}\,\frac{dx^\nu}{d\tau}\frac{dx^\lambda}{d\tau}
	+ \frac{q}{m}\,F^i{}_{\ \nu}\,\frac{dx^\nu}{d\tau}, \\
	\frac{d^2 t}{d\tau^2} &= -\Gamma^0_{\nu\lambda}\,\frac{dx^\nu}{d\tau}\frac{dx^\lambda}{d\tau}
	+ \frac{q}{m}\,F^0{}_{\ \nu}\,\frac{dx^\nu}{d\tau}.
\end{split}
\end{align}

Substituting $\dfrac{dx^\mu}{d\tau}=\dfrac{dt}{d\tau}\,\dfrac{dx^\mu}{dt}$ and simplifying, we obtain the equation of motion in terms of coordinate time. Writing explicitly with $v^i\equiv dx^i/dt$, one finds
\begin{align}
	\label{Roth}
	\frac{d^2 x^i}{dt^2} &=
	-\Gamma^i_{00}
	-2\,\Gamma^i_{0j}\,v^j
	-\Gamma^i_{jk}\,v^j v^k \notag \\
	&\quad +\Big[\Gamma^0_{00}
	+2\,\Gamma^0_{0j}\,v^j
	+\Gamma^0_{jk}\,v^j v^k\Big]\,v^i \notag \\
	&\quad + \frac{q}{m}\frac{d\tau}{dt}\Big(F^i{}_{\ 0}+F^i{}_{\ j}v^j - F^0{}_{\ j}v^j\,v^i\Big).
\end{align}
This equation defines a semispray with coefficients $G^i(x,v)$. The derivation of this relation follows the treatment in Sec.~9.1 of Weinberg’s text~\cite{Weinberg}, with the extension here to include the electromagnetic interaction.

\subsubsection{Static spacetime}

Now we consider the static spacetime with $g_{0i}=0$ and denote $g_{00}=-V^2(x)$, that is
\begin{align}
	ds^2=-V^2(x)\,dt^2+g_{ij}(x)\,dx^i dx^j,\quad i,j=1,2,3.
\end{align}

In this background, the nonvanishing Christoffel symbols are
\begin{align}
	\begin{split}
		\label{Whitman}
		\Gamma^{0}{}_{0i}&=\tfrac12\,\partial_i\ln V^2,\\
		\Gamma^{i}{}_{00}&=\tfrac12\,\partial^{i}V^2,\\
		\Gamma^{i}{}_{jk}&=\Gamma^{i}{}_{jk}[g].
	\end{split}
\end{align}
Here, $\Gamma^{i}{}_{jk}[g]$ denotes the Christoffel symbols associated with the spatial metric $g$. For brevity, the symbol $[g]$ will be omitted in what follows.

In addition, the corresponding nonvanishing components of the electromagnetic field tensor are
\begin{align}
	\label{Eliot}
	\begin{split}
		F^{i}{}_{0}&=g^{ik}\partial_k A_0,\\
		F^{i}{}_{j}&=g^{ik}(A_{j,k}-A_{k,j}),\\
		F^{0}{}_{j}&=\frac{1}{V^2}\partial_j A_0
	\end{split}
\end{align}

The conserved energy associated with the timelike Killing field $\partial_t$ is
\begin{align}
	E \,\equiv\, -\!\left(m\,g_{0\mu}\frac{dx^\mu}{d\tau}+q\,A_0\right)
	= m\,V^2\,\frac{dt}{d\tau}-q\,A_0.
\end{align}
It follows that
\begin{align}
	\label{Stevens}
	\frac{dt}{d\tau}=\frac{E+qA_0}{mV^2}.
\end{align}

Substituting Eqs.~\eqref{Whitman},~\eqref{Eliot} and~\eqref{Stevens} into Eq.~\eqref{Roth} yields: 
\begin{align}
	\label{Miłosz}
		\frac{d^2 x^i}{dt^2}
		&+\tfrac12\,\partial^{i}V^2
		+\Gamma^{i}{}_{jk}\,v^j v^k
		-\big(\partial_j\!\ln V^2\big)\,v^j v^i\nn\\
		&=\frac{q V^2}{E+qA_0}\!\left[g^{ik}\partial_k A_0+g^{ik}(\partial_kA_j-\partial_jA_k)\,v^j\right]\nn\\
		&~~~~~-\frac{q}{E+qA_0}\,\partial_j A_0\,v^j\,v^i,
\end{align}
which may be regarded as the geodesic equation of a semispray with coefficients $G^i(x,v)$, i.e.,
\begin{equation}
	\frac{d^2 x^i}{dt^2} + 2G^i(x,v) = 0 ,
\end{equation}
with $G^i(x,\lambda v)\neq \lambda^2 G^i(x,v)$.


\section{Optical Metric} 
\label{Hamilton}

We begin with the simple case of massless particles, which already encapsulates the essence of our method. 

For a neutral particle ($q=0$), Eq.~\eqref{Miłosz} reduces to
\begin{align}
	\label{Szymborska}
	\frac{d^2 x^i}{dt^2}
	+\Gamma^{i}{}_{jk}v^j v^k
	-\big(\partial_j\ln V^2\big)v^j v^i+\tfrac12\,\partial^{i}V^2=0.
\end{align}
For a null ray, the constraint is
\begin{equation}
	\label{Leibniz}
	g(x,v)=g_{ij}v^iv^j=V^2.
\end{equation}
Accordingly, terms independent of $v$ are rewritten as quadratic ones by using the null constraint, namely
\begin{equation}
	\frac12\partial^{i}\!(V^2)\to\frac12\partial^{i}(V^2)\frac{g_{kj}v^kv^j}{V^2}
	=\frac12\partial^{i}\big(\ln V^2\big)g_{kj}v^kv^j.
\end{equation}
With this replacement, Eq.~\eqref{Szymborska} becomes
\begin{align}
	\label{Lagrange}
	\frac{d^2 x^i}{dt^2} + 2\,G^i(x,v) = 0,
\end{align}
with spray coefficients
\begin{align}
	G^i(x,v) =& \tfrac12\Big\{\Gamma^{i}{}_{jk}
	+\tfrac12\big[\delta^i{}_j\,\partial_k\!\ln \tfrac{1}{V^2}
	+\delta^i{}_k\,\partial_j\!\ln\tfrac{1}{V^2}\nn\\
	&- g_{jk}g^{il}\,\partial_l\!\ln \tfrac{1}{V^2}\big]\Big\}\,v^j v^k.
\end{align}
This defines a Riemannian spray. From Eqs.~\eqref{Zagajewski}--\eqref{Ernaux}, it follows that it is conformally related to $g$ with factor 
$\Phi(x)=1/V^2$:
\begin{equation}
	h(x,y) = \frac{1}{V^2} g(x,y).
\end{equation}

Thus, the coordinate time $t$ serves as an affine parameter along the spatial projection of null geodesics, i.e. light trajectories. This is precisely Fermat’s principle: the physical path of light extremizes the travel time. The corresponding metric $h$, referred to as the \emph{optical metric}, has the line element
\begin{equation}
	dt^2 = \frac{1}{V^2}\, g_{ij}\,dx^i dx^j .
\end{equation}

Although the constraint~\eqref{Leibniz} already suggests this form, it is only through Eq.~\eqref{Lagrange}—or, in the standard treatment, through Fermat’s principle—that one confirms light rays are indeed geodesics of the optical metric.


\section{Jacobi Metric for Massive Particles}
\label{Einstein}

To proceed beyond the massless case, we consider timelike particles with rest mass $m$ and conserved energy $E$. The normalization of the four-velocity reads
\begin{equation}
	-\,V^2\!+g_{ij}\,v^iv^j=-\left(\frac{d\tau}{dt}\right)^2.
\end{equation}
The conserved energy is
\begin{equation}
	E=m\,V^2\,\frac{dt}{d\tau}.
\end{equation}
Eliminating $d\tau/dt$ yields the quadratic constraint  
\begin{equation}
	\label{Cartier-Bresson}
	g(x,v)=g_{ij}\,v^i v^j = V^2\left(1-\frac{m^2 V^2}{E^2}\right).
\end{equation}
Using this constraint, the last term in Eq.~\eqref{Szymborska} can be rewritten as
\begin{align}
	\frac12\,\partial^i V^2
	&\to\left(\frac12\,\partial^i V^2\right)\,
	\frac{g_{kl}\,v^k v^l}{V^2\left(1-\tfrac{m^2 V^2}{E^2}\right)} \nn\\
	&=\frac12\,g_{kl}\,\partial^i\ln\!\Big(\frac{V^2}{E^2-m^2V^2}\Big)v^k v^l.
\end{align}

Accordingly, the spray coefficients take the form
\begin{align}
	G^i(x,v) =& \tfrac12\Big\{\Gamma^{i}{}_{jk}
	+\tfrac12\big[\delta^i{}_j\,\partial_k\!\ln \tfrac{1}{V^2}
	+\delta^i{}_k\,\partial_j\!\ln\tfrac{1}{V^2} \nn\\
	&\hspace{0.5cm}- g_{jk}g^{il}\,\partial_l\ln\!\Big(\tfrac{E^2-m^2V^2}{V^2}\Big)\big]\Big\}v^j v^k.
\end{align}

The above expression defines a spray of the Berwald type. Nevertheless, identifying the corresponding metric is nontrivial. To address this, we introduce a new parameter $\ell$ along the trajectory, and denote
\begin{align}
	\label{Duras}
	\begin{split}
	f(x)&\equiv \frac{dt}{d\ell}, \quad
	y^i\equiv \frac{dx^i}{d\ell}, \\
	v^i&=\frac{dx^i}{dt}=\frac{y^i}{f},\quad f'\equiv \frac{df}{d\ell}.
	\end{split}
\end{align}
This parametrization enables us to reformulate the dynamics entirely in terms of spatial quantities.

By applying the chain rule, the second derivative with respect to the coordinate time $t$ can be written as
\begin{align}
	\label{Ondaatje}
	\frac{d^2x^i}{dt^2}
	=\frac{1}{f^2}\left(\frac{d^2x^i}{d\ell^2}-\frac{f'}{f}\,y^i\right).
\end{align}

Accordingly, the $t$-parameterized equation can be reformulated in terms of the $\ell$-parameter as
\begin{align}
	\label{Balzac}
	\frac{d^2x^i}{d\ell^{2}}
	&+\bigg[\Gamma^{i}{}_{jk}
	-\frac12\,g_{jk}\,g^{il}\,\partial_l\ln\Phi(x)\bigg]y^j y^k\nn\\
	&=\partial_k \ln (V^2 f)\,y^k y^i,
\end{align}
where we have defined
\begin{equation}
	\Phi(x)\equiv \frac{E^2-m^2 V^2}{V^2}.
\end{equation}

Next, we add the following term to both sides of Eq.~\eqref{Balzac}:  
\begin{equation}
	\frac12 \left[\delta^i{}_j\,\partial_k\!\ln \Phi(x)
	+\delta^i{}_k\,\partial_j\!\ln \Phi(x)\right] y^j y^k .
\end{equation}
As a result, we obtain
\begin{align}
	\label{Modiano}
	\frac{d^2x^i}{d\ell^{2}}
	+2\tilde{G}^{i}
	=\partial_k \ln \!\big(V^2 f \Phi\big)\,y^k y^i ,
\end{align}
where the modified spray coefficients take the form
\begin{align}
	\label{Stendhal}
	\tilde{G}^{i}
	=G_g^{i}
	&+\tfrac14\Big(\delta^i{}_j\,\partial_k\!\ln \Phi
	+\delta^i{}_k\,\partial_j\!\ln \Phi
	\nn\\
	&-g_{jk}\,g^{il}\,\partial_l\!\ln \Phi\Big)\,y^j y^k .
\end{align}

Comparing with Eq.~\eqref{Perec}, we recognize that $\tilde{G}^i$ coincides with the spray coefficients associated with the conformally rescaled Riemannian metric
\begin{equation}
	h(x,y)=\Phi(x)\,g(x,y)
	=\frac{E^2-m^2V^2}{V^2}\,g(x,y).
\end{equation}

If we further choose the parameter $\ell$ such that
\begin{equation}
	f=\frac{dt}{d\ell}=\frac{C}{V^2\Phi}
	=\frac{C}{E^2-m^2V^2},
\end{equation}
with $C$ a constant, then Eq.~\eqref{Modiano} reduces precisely to the geodesic equation of the metric $h$. 
In this parametrization, $\ell$ is the arc-length parameter of the metric
\begin{equation}
	\label{Chekhov}
	d\ell^2 
	= \frac{E^2 - m^2 V^2}{V^2}\, g_{ij}\, dx^i dx^j .
\end{equation}
This is precisely the \emph{Jacobi metric} for a particle of rest mass $m$ and conserved energy $E$ in a static spacetime. 
It was first introduced by Gibbons on the basis of the Maupertuis--Jacobi principle~\cite{Gibbons2016}.

The constraint~\eqref{Cartier-Bresson} directly implies the following time metric,  
\begin{equation}
	\label{Stravinsky}
	dt^2 \;=\;
	\frac{E^2}{V^2\!\left(E^2-m^2 V^2\right)}\,g_{ij}\,dx^i dx^j .
\end{equation}
One may regard this as an ``optical metric'' for massive particles.  
Although the orbits of massive particles are not geodesics in this effective space, the metric can nevertheless be useful for studying phenomena such as the time delay of particle trajectories.

From Eqs.~\eqref{Chekhov} and~\eqref{Stravinsky} we infer that the constant is fixed as $C=E$.  
Returning to Eq.~\eqref{Stendhal}, we note that the conformal factor $\Phi(x)$ is not unique: it is defined only up to a positive multiplicative constant, i.e.,
\begin{equation}
	\Phi(x) \;\to\; C_0\,\Phi(x), \qquad C_0>0.
\end{equation}
Such a rescaling does not alter the associated spray or the resulting Jacobi metric. The specific choice $C_0 = 1/E^{2}$ was introduced by Crisnejo and Gallo~\cite{Crisnejo-Gallo}, derived through a correspondence between the motion of a massive particle and the dynamics of light rays in a homogeneous plasma.


\section{Jacobi Metric for Charged Particles}
\label{Maxwell}

We now generalize the Jacobi metric construction to describe the motion of a charged particle in a static spacetime endowed with an electromagnetic field, whose associated semispray is given by Eq.~\eqref{Miłosz}.

The normalization of the four-velocity takes the form
\begin{align}
	-\,V^2 + g_{ij}v^i v^j = -\left(\frac{d\tau}{dt}\right)^2 .
\end{align}
Meanwhile, the conserved energy of a particle with rest mass $m$ and charge $q$ is expressed as
\begin{align}
	E = m\,V^2\,\frac{dt}{d\tau} - q A_0 .
\end{align}
Combining these two relations yields the quadratic spatial constraint
\begin{align}
	\label{Kafka}
	g(x,v)=g_{ij}v^i v^j
	= V^2\!\left(1 - \frac{m^2 V^2}{(E+qA_0)^2}\right) .
\end{align}

Next, by combining the $v$-independent terms in Eq.~\eqref{Miłosz} 
and invoking the constraint~\eqref{Kafka}, we obtain
\begin{align}
	\label{Proust}
	&\tfrac12\,\partial^{i}(V^2)
	-\frac{q\,V^2}{E+qA_0}\,g^{ik}\partial_k A_0
	\nn\\
	&\quad\to\;
	\tfrac12\,\big(g_{kl}v^k v^l\big)\,
	\partial^{i}\!\ln\!\Bigg(
	\frac{V^2}{(E+qA_0)^2 - m^2 V^2}\Bigg).
\end{align}

The electromagnetic contribution involving the spatial field tensor $F^i{}_j$ 
can likewise be recast into a form homogeneous of degree two in the velocities,
\begin{align}
	\label{Hemingway}
	\frac{q F^i{}_j\,V^2\,v^j}{E+qA_0}
	\;\to\;
	qF^i{}_j\,v^j\;
	\sqrt{\;\frac{\,V^2\,g(x,v)\,}{\,(E+qA_0)^2 - m^2 V^2\,}\;}.
\end{align}

In addition, the pair of quadratic terms in the velocities can be combined as
\begin{align}
	\label{Mauriac}
	&-\big(\partial_j\!\ln V^2\big)\,v^j v^i
	+\frac{q}{E+qA_0}\,(\partial_j A_0\,v^j)\,v^i
	\nn\\
	&\qquad=\;
	-\partial_j\!\ln\!\Bigg(\frac{V^2}{E+qA_0}\Bigg)\,v^j v^i .
\end{align}

Substituting Eqs.~\eqref{Proust}, \eqref{Hemingway}, and \eqref{Mauriac} into Eq.~\eqref{Miłosz}, we obtain
\begin{align}
	\label{Hilbert}
	&0=\frac{d^2x^i}{dt^{2}}
	+ \Gamma^{i}{}_{jk}\,v^j v^k
	-\tfrac12\,g(x,v)\,\partial^{i}\!\ln\!\Phi_q(x) \nn\\
	&
	-\partial_{j}\!\ln\!\Bigg(\frac{V^2}{E+qA_0}\Bigg)\,v^j v^i
	- q\,\sqrt{\frac{g(x,v)}{\Phi_q(x)}}\,F^i{}_j\,v^j,
\end{align}
with
\begin{align}
	\Phi_q(x) = \frac{(E+qA_0)^2 - m^2 V^2}{V^2} .
\end{align}

Equation~\eqref{Hilbert} defines a spray, but not of Berwald type, except in the special cases $q=0$ or $F_{ij}=0$. As no metric can be directly reconstructed from this spray, we proceed as in the neutral massive case by introducing a new parameter~$\ell$ and adopting the notation of Eq.~\eqref{Duras}.

Using Eq.~\eqref{Ondaatje}, the $t$-parameter equation~\eqref{Hilbert} can be rewritten in the $\ell$-parameter form
\begin{align}
	&~~~0 = \frac{d^2x^i}{d\ell^{2}}
	+ \Gamma^{i}{}_{jk}\,y^j y^k
	-\tfrac12\,g(x,y)\,\partial^{i}\!\ln\!\Phi_q(x) \nn\\
	&\quad
	-\partial_{j}\!\ln\!\Bigg(\frac{fV^2}{E+qA_0}\Bigg)\,y^j y^i
	- q\,\sqrt{\frac{g(x,y)}{\Phi_q(x)}}\,F^i{}_j\,y^j .
\end{align}
This can further be expressed in conformal form as
\begin{align}
	&\frac{d^2x^i}{d\ell^{2}}
	+\bigg\{\Gamma^{i}{}_{jk}+\tfrac12\bigg[\delta^i{}_j\,\partial_k\!\ln \Phi_q
	+\delta^i{}_k\,\partial_j\!\ln \Phi_q\nn\\
	&~~~-\,g_{jk}\,
	g^{il}\partial_{l}\!\ln\!\Phi_q\bigg]\bigg\}\,y^j y^k
	- q\,\sqrt{\frac{g(x,y)}{\Phi_q}}\,F^i{}_j\,y^j\nn\\
	&~~~-\partial_{j}\!\ln\!\Bigg(\frac{fV^2\Phi_q}{E+qA_0}\Bigg)\,y^j y^i = 0 .
\end{align}

We set
\begin{align}
	\frac{f V^2 \Phi_q}{E+qA_0}
	= \frac{f\big[(E+qA_0)^2 - m^2 V^2\big]}{E+qA_0}
	= C ,
\end{align}
or equivalently,
\begin{align}
	f \;=\; \frac{dt}{d\ell}
	= \frac{(E+qA_0)\,C}{(E+qA_0)^2 - m^2 V^2} .
\end{align}
The equation can then be cast in the standard spray form
\begin{align}
	\frac{d^2x^i}{d\ell^{2}} + 2G^i = 0 ,
\end{align}
with
\begin{align}
	\label{Rilke}
	G^i
	= G^i_\alpha
	+ \frac12 q\,\alpha(y)\,a^{ij}\left(A_{j,k}-A_{k,j}\right)\,y^k ,
\end{align}
where
\begin{align}
	G_\alpha^i
	= &G_g^i
	+ \frac14\!\bigg[
	\delta^{i}{}_{j}\,\partial_k\ln \Phi_q
	+ \delta^{i}{}_{k}\,\partial_j\ln \Phi_q\nn\\
	&~~~~~~~- g_{jk}\,g^{il}\,\partial_{l}\ln \Phi_q
	\bigg] y^j y^k
\end{align}
is the spray associated with the metric 
\begin{align}
	\alpha(x,y) = \sqrt{a_{ij}y^i y^j}, \qquad 
	a_{ij} = \Phi_q\, g_{ij}.
\end{align}

By comparison with Eq.~\eqref{Russell}, we confirm that the spray~\eqref{Rilke} corresponds to a Randers-type Finsler metric of the form
\begin{align}
	F(x,dx) = \alpha(x,dx) + \beta(x,dx) ,
\end{align}
where
\begin{align}
	\alpha(x,dx) &= \sqrt{\frac{(E+qA_0)^2 - m^2 V^2}{V^2}\,g_{ij}\,dx^i dx^j}, \nn\\
	\beta(x,dx)  &= q A_i dx^i ,
\end{align}
with the requirement $a^{ij}b_ib_j<1$. This Randers metric aligns with Chanda's general construction for stationary spacetimes under the condition $ g_{0i} = 0 $. Later, Li \emph{et al.} applied it to investigate the lensing of charged particles in a Schwarzschild spacetime endowed with a dipole magnetic field~\cite{massiveGB-LiWJ}. If the magnetic potential vanishes ($ A_i = 0 $), the $ \beta $-term disappears, reducing the structure to a purely Riemannian metric. This Riemannian case was first used by Das \emph{et al.} to analyze charged particle dynamics in the Reissner--Nordström spacetime~\cite{Das-EPJC-2017}.

It should be noted that $\ell$ is not the length parameter of the Randers metric, but rather the length parameter of its Riemannian part, namely
\begin{align}
	d\ell^2 =\alpha^2(x,dx)=a_{ij}\,dx^i dx^j.
\end{align}
Accordingly, in conjunction with the quadratic constraint~\eqref{Kafka}, one finds $C=1$.


\section{Jacobi Metric for the Planar Circular Restricted Three-Body Problem}
\label{Poincaré}

Consider the Planar Circular Restricted Three-Body Problem (PCR3BP) in a uniformly rotating frame. Two primaries, with total mass normalized to unity, are fixed at the positions
\((-\mu,0)\) and \((1-\mu,0)\), where \(\mu\in(0,1)\) denotes the mass parameter. The infinitesimal mass is located at \((x^1,x^2)=(x,y)\). 
With the standard non-dimensionalization, the effective potential is
\begin{align}
	U(x^1,x^2)=\frac{1-\mu}{r_1}+\frac{\mu}{r_2}+\tfrac12\big((x^1)^2+(x^2)^2\big),
\end{align}
where
\begin{align}
\begin{split}
	&r_1=\sqrt{(x^1+\mu)^2+(x^2)^2},\\
	&r_2=\sqrt{(x^1-1+\mu)^2+(x^2)^2}.
\end{split}
\end{align}
The equations of motion in the rotating frame are
\begin{align}
	\label{Wittgenstein}
	\begin{split}
	\frac{d^2 x^1}{dt^2} - 2\,v^2 &= \partial_{x^1}U, \\
	\frac{d^2 x^2}{dt^2} + 2\,v^1 &= \partial_{x^2}U,
	\end{split}
\end{align}
where we denote $v^i= \tfrac{dx^i}{dt}$.

The system possesses the Jacobi integral, denoted by \(C_J\),
\begin{equation}
	C_J = 2\,U(x) - \delta_{ij}v^i v^j ,
\end{equation}
which may be equivalently expressed as the quadratic constraint
\begin{align}
	\label{Kant}
\begin{split}
	&\delta(x,v)=\delta_{ij}v^i v^j = \Phi(x), \\
	&\Phi(x) := 2U(x) - C_J .
\end{split}
\end{align}
The accessible region of motion is thus characterized by the condition \(\Phi(x) > 0\).  

The equations of motion~\eqref{Wittgenstein} can be rewritten as
\begin{align}
	\frac{d^2 x^i}{dt^2}- \delta^{ij}\,\partial_j U- 2\,\delta^{ij}\,\epsilon_{jk}\,v^k= 0 .
\end{align}
where $\epsilon_{ij}$ denotes the two-dimensional antisymmetric symbol,
\begin{align}
	\epsilon_{ij}=
	\begin{pmatrix}
		0 & 1 \\
		-1 & 0
	\end{pmatrix}.
\end{align}

Using the constraint~\eqref{Kant}, the equation of motion becomes
\begin{align}
	\label{Hawking}
	0
	&= \frac{d^2 x^i}{dt^2}
	- \delta^{ij}\partial_j U \,\frac{\delta(x,v)}{\Phi(x)}
	- 2\,\delta^{ij}\,\epsilon_{jk}\,v^k
	\sqrt{\frac{\delta(x,v)}{\Phi(x)}} \nn\\
	&= \frac{d^2 x^i}{dt^2}
	- \frac12\,\delta^{ij}\partial_j\!\ln\Phi(x)\;
	\delta(x,v)\nn\\
	&~~~~~- 2\,\delta^{ij}\,\epsilon_{jk}\,v^k
	\sqrt{\frac{\delta(x,v)}{\Phi(x)}} \nn\\
	&= \frac{d^2 x^i}{dt^2} + 2G_\alpha^i
	- 2\,a^{ij}\epsilon_{jk}\,\alpha(x,v)v^k
	\nn\\
	&~~~~~- \partial_k\!\ln\Phi(x)\; v^k v^i ,
\end{align}
where
\begin{align}
	G_\alpha^i
	=& \frac14\Big[
	\delta^{i}{}_{j}\,\partial_k\!\ln \Phi(x)
	+ \delta^{i}{}_{k}\,\partial_j\!\ln \Phi(x)\nn\\
	&- \delta_{jk}\,\delta^{il}\,\partial_{l}\!\ln \Phi(x)
	\Big] v^j v^k,
\end{align}
which are the spray coefficients of the Riemannian metric
\begin{equation}
	\alpha(x,v)=\sqrt{a_{ij}v^iv^j},\quad a_{ij}=\Phi(x)\delta_{ij}.
\end{equation}
Introducing a new parameter $\ell$, we define
\begin{equation}
	y^i = \frac{dx^i}{d\ell}, 
	\qquad 
	f = \frac{dt}{d\ell},
\end{equation}
and choose
\begin{equation}
	\label{Landau}
	f = \frac{C_0}{\Phi(x)} .
\end{equation}
With this choice, the equations of motion~\eqref{Hawking} take the form
\begin{equation}
	\label{Bonnefoy}
	\frac{d^2 x^i}{d\ell^{2}}
	+2G^i_\alpha
	+\,a^{ij}\,(b_{j,k}-b_{k,j})\,\alpha(y)\,y^k = 0,
\end{equation}
where $b_i$ satisfies
\begin{equation}
	b_{j,k}-b_{k,j}=2\,\epsilon_{kj}, 
	\qquad \epsilon_{12}=+1 .
\end{equation}

A convenient choice (unique up to a gauge $b_i\!\to b_i+\partial_i\lambda$) is
\begin{equation}
	b_1=-x^2,\qquad b_2=x^1.
\end{equation}

By comparison with Eq.~\eqref{Russell}, we confirm that the spray from Eq.~\eqref{Bonnefoy} corresponds to a Randers-type Finsler metric of the form
\begin{align}
	\label{Feynman}
	F(x,dx) = \alpha(x,dx) + \beta(x,dx),
\end{align}
where
\begin{align}
\begin{split}
	\alpha(x,dx) &= \sqrt{\left[2U(x)-C_J\right]\,\delta_{ij}\,dx^i dx^j},\\
	\beta(x,dx)  &=x^1\,dx^2-x^2\,dx^1.
\end{split}
\end{align}
Note that $\ell$ is the length parameter of the Riemannian metric $\alpha$, and the integration constant in Eq.~\eqref{Landau} is fixed as $C_0=1$.

For $F$ to define a strongly convex Finsler structure, one requires
\begin{equation}
	\|\beta\|_\alpha^2
	= a^{ij} b_i b_j < 1 ,
\end{equation}
which in the present case amounts to
\begin{equation}
	\frac{(x^1)^2+(x^2)^2}{2U(x)-C_J} < 1.
\end{equation}

The Randers metric is well defined only in the domain where $\Phi(x)=2U(x)-C_J>0$, i.e.\ inside the Hill's region determined by the zero-velocity curve. Moreover, the strong convexity condition requires 
	$\|\beta\|_\alpha<1$, which is equivalent to $\Phi(x)=2U(x)-C_J>(x^1)^2+(x^2)^2$.
Hence the strictly valid Finsler structure is confined to the subregion where $\Phi(x)$ dominates over the radial distance squared. In particular, at the boundary $\Phi(x)=0$ the metric degenerates, and the effective geometric description breaks down.

It is of considerable interest to study this problem within the framework of Finsler geometry, or to reformulate it in terms of the Zermelo navigation problem. Exploring the geometric structures of more complex three-body problems would also be interesting, but such investigations lie beyond the scope of the present work.


\section{conclusion}
\label{Li Zonghai}

This paper develops a systematic procedure for geometrizing dynamics. For dynamical systems constrained by the quadratic relation~\eqref{Yourcenar}, the associated semispray can be elevated to a spray, from which a metric structure is extracted through an appropriate reparameterization. In this way, without invoking Fermat’s principle or the Maupertuis–Jacobi principle, we reconstruct in static spacetimes the optical metric, the Jacobi metric for massive particles, and the Jacobi metric for charged particles in electromagnetic fields. Moreover, we show that the planar circular restricted three-body problem naturally leads to a Randers–Finsler metric.

Regarding the characteristics and potential of this method, a few remarks are in order. First, it establishes a unified framework for extracting metric structures from differential equations and their constraints, a process independent of the specific physical objects or scenarios under consideration. This universality gives it the potential to extend beyond the scope of traditional mechanics to broader disciplinary fields, for instance to constrained dynamical systems typically studied within the KCC framework. Second, it simplifies problem formulation: the apparent complexity of semisprays and sprays may arise from inconvenient parameter choices, yet through reparameterization they can be endowed with a suitable metric form, thereby transforming complex dynamical problems into intuitive geometric ones in a metric space, centered on the study of geodesics and curvature. Third, it introduces a “geometry-first” perspective: the derivation proceeds independently of the traditional action principle, taking the geodesic equation directly as the starting point to uncover the underlying metric structure. This viewpoint offers a promising direction for the study of geometric dynamics, with potential applications in perturbative analysis and control problems. 

In addition, this work provides concrete examples for the abstract study of spray metrizability. Conversely, invoking appropriate mathematical theorems may enhance the operational effectiveness of the present method. The examples presented in this work are based on the quadratic constraint~\eqref{Yourcenar} and utilize a conformal Riemannian metric structure. Under the same conditions, many interesting metrics may be identified. The methodology developed here may also prove valuable for studying systems governed by the more general constraint~\eqref{Plato}, though such an extension requires familiarity with the spray formulation of specific Finsler metrics. 


\end{document}